\documentclass[pdflatex,sn-nature]{sn-jnl}

\usepackage{graphicx}%
\usepackage{multirow}%
\usepackage{amsmath,amssymb,amsfonts}%
\usepackage{amsthm}%
\usepackage{mathrsfs}%
\usepackage[title]{appendix}%
\usepackage{xcolor}%
\usepackage{textcomp}%
\usepackage{manyfoot}%
\usepackage{booktabs}%
\usepackage{algorithm}%
\usepackage{algorithmicx}%
\usepackage{algpseudocode}%
\usepackage{listings}%

\usepackage[normalem]{ulem}

\usepackage{lineno}

\theoremstyle{thmstyleone}%
\theoremstyle{thmstyletwo}%

\theoremstyle{thmstylethree}%

\begin{document}
\title[Article Title]{Neuron Surface Emitting Laser (NeuronSEL): Spiking Regimes and Negative Differential Resistance in Solitary Multi-junction VCSELs}


\author*[1]{\fnm{Maria} \sur{Duque-Gijon}}\email{maria.duque-gijon@strath.ac.uk}

\author*[1]{\fnm{Joshua} \sur{Robertson}}\email{joshua.robertson@strath.ac.uk}

\author[1]{\fnm{Dafydd} \sur{Owen-Newns}}\email{dafydd.owen-newns@strath.ac.uk}

\author[2]{\fnm{Jack} \sur{Baker}}\email{bakerj19@cardiff.ac.uk}

\author[2]{\fnm{Craig P.} \sur{Allford}}\email{allfordcp1@cardiff.ac.uk}

\author[1]{\fnm{Xavier} \sur{Porte}}\email{javier.porte-parera@strath.ac.uk}

\author[2]{\fnm{Samuel} \sur{Shutts}}\email{shuttss@cardiff.ac.uk}

\author[2]{\fnm{Peter M.} \sur{Smowton}}\email{smowtonpm@cardiff.ac.uk}

\author[1]{\fnm{Antonio} \sur{Hurtado}}\email{antonio.hurtado@strath.ac.uk}



\affil*[1]{\orgdiv{Institute of Photonics, SUPA, Dept. of Physics}, \orgname{University of Strathclyde}, \orgaddress{\street{16 Richmond St}, \city{Glasgow}, \postcode{G1 1XQ}, \state{Glasgow}, \country{UK}}}

\affil[2]{\orgdiv{Compound Semiconductor Technology Group}, \orgname{Cardiff University}, \orgaddress{\street{Translational Research Hub}, \city{Cardiff}, \postcode{CF24 4HQ}, \state{Wales}, \country{UK}}}

\abstract{Neuromorphic photonics is emerging as a powerful platform for fast and efficient optical information processing and sensing. However, future brain-inspired photonic systems require compact and scalable light sources, capable of generating the neuro-mimetic optical signals needed for their operation. This work demonstrates a single-stack laser that delivers optical and electrical neural-like spiking emission under solitary operation. Termed the Neuron Surface-Emitting Laser (NeuronSEL), this compact, multi-junction Vertical-Cavity Surface Emitting Laser (VCSEL) exhibits non-linear Negative Differential Resistance (NDR), similar to that observed in memristive devices. Leveraging this NDR behaviour enables the novel demonstration of multiple neuronal features in the NeuronSEL including refractoriness and threshold-/integrate-and-fire dynamics. We demonstrate the NeuronSEL’s behaviour as an optical spiking neuron and its ability to perform processing functions, such as coincidence detection and exclusive OR operations. Its scalability is illustrated by proposing a network based on an array of NeuronSELs, able to perform classification tasks. The NeuronSEL emerges as a strong candidate for practical and scalable neuromorphic photonic hardware, with potential impact across a range of applications in optical sensing, communications and computing technologies, whilst benefitting from the inherent advantages of VCSEL technology —low manufacturing cost, compactness, efficiency, vertical emission, and straightforward integration into large arrayed-structures and networks.}




\maketitle

\section{Introduction}\label{Introduction}
Neuromorphic photonics, a field that emulates and deploys the principles of brain-inspired computation and the unique properties of light-based hardware, is creating new pathways towards high-speed and energy-efficient information processing and sensing~\cite{markovic2020physics, christensen20222022}. Recently, interest in neuromorphic photonics has increased due to the growing demand for computationally-efficient, low-power processing systems, driven primarily by the boom of artificial intelligence~\cite{shastri2021photonics,brunner2025roadmap}. Photonic platforms inherently offer ultrafast response times and large bandwidths, motivating the development of hardware photonic neuronal models~\cite{Xiang2021}, and photonic neural networks~\cite{Ashtiani2022,Giamougiannis2023,Biasi2024,Tossoun2025} that operate multiple orders of magnitude faster than biological systems. Spiking Neural Networks (SNNs), the bio-realistic approach to neural networks that make use of event-driven signalling mechanisms, have also been gaining interest on the photonic platform~\cite{Feldmann2019,eshraghian2023training,owen2025photonic}
. Inspired by their biological counterparts, SNNs communicate using discrete spikes, encoding information in sparse signals that significantly reduce energy consumption and provide inherent temporal processing capabilities attractive for ultra-fast, low-power neuromorphic hardware. Spiking-enabled hardware systems have found use in sensory functionalities with applications including dynamic vision and event-driven auditory sensors, and more recently, computational photonic and electronic sensors that generate spike trains directly at the pixel level
~\cite{oconnor2013real,zhou2023computational}. In parallel, neuromorphic platforms integrating sensing and computation are highly desirable for novel applications such as edge-computing and smart-sensors, given their potentials for real-time, low-energy, information processing without storing or transferring large amounts of data~\cite{Zhou2020}.

In recent years, photonic approaches have deployed a variety of elements, e.g. microring resonators (MRRs)
~\cite{Biasi2024pnns,Borghi2021,Donati2025all-optical}, phase change materials~\cite{rios2019memory,feldmann2021parallel} and modulators~\cite{xu2019high,Prabhu2020,Pai2023}, among others~\cite{Wang2025integrated}, to implement processing functionalities (e.g. optical weighting, nonlinear activation functions, etc.) for optical neural networks. However, despite this rapid progress, only a minority of works have focused on the implementation of bio-realistic photonic systems, including approaches based on semiconductor lasers~\cite{Xiang2024semicon,Huang2026}, MRRs~\cite{Donati2025all-optical}, and resonant tunnelling diodes (RTDs)~\cite{Donati2025,OwenNewns2026,hejda2023artificial}, to name a few. For instance, photo-detecting RTDs have demonstrated that their inherent resonant quantum tunnelling effects permit the achievement of non-linear regions of negative differential resistance (NDR) and excitability, which can be exploited to obtain controllable optically-triggered neural-like spiking~\cite{hejda2023artificial,Donati2025,OwenNewns2026}. Similarly, NDR regions are also often displayed by memristors, profusely researched electronic devices for neuromorphic hardware platforms~\cite{Mehonic2020}. Coined the missing 4$^{th}$ fundamental circuit element, memristors have demonstrated regions of NDR through either voltage- or current-controlled mechanisms (e.g filament formation), to achieve operation as both synaptic elements with non-volatile memory and spiking neuron circuits~\cite{Kumar2017,peng2024memristor,Pei2025}
Vertical-Cavity Surface-Emitting Lasers (VCSELs) are another stand out candidate for neuromorphic photonics given their wide deployment, low fabrication cost and compact structure~\cite{skalli2022photonic}. VCSELs have characteristics well aligned with the goal of efficient and scalable photonic systems offering coherent light emission with low threshold currents, high wall-plug efficiency, ease of integration into large 2D arrays, and high-coupling efficiency to optical fibres for seamless compatibility with optical communication networks~\cite{haghighi2021electrically,Pan2024Harness}. These features make them especially attractive for deployment in large, compact, spatial architectures~\cite{Chen2022} suitable for scalable optical neural networks and sensing platforms. Multi-junction VCSELs, a variant of traditional VCSELs that stack active light emitting junctions, are also highly interesting for sensing applications~\cite{Staudinger2023}. The high output power achieved by these devices is rising to challenge traditional edge-emitting lasers in LiDAR, and their high-power conversion efficiency makes them candidates for future implementation in energy-conscious data centres~\cite{Staudinger2023, Zhou2024Thermal,Xiao2024}. The potential deployment of VCSELs for neuromorphic operations has rapidly developed from its initial postulation (see~\cite{hurtado2015controllable}), to systems of single VCSEL-based artificial neurons~\cite{Turconi2013,robertson2020ultrafast,Pammi2020}, and more recently into demonstrations of photonic reservoir computers~\cite{Vatin2019,bueno2021comprehensive,Porte2021} and time-multiplexed photonic SNNs
~\cite{Owen_GHzRate2023,owen2025photonic}. However, despite the exciting direction of VCSELs for neuromorphic photonics, these implementations still rely on complex mechanisms such as external optical injection and optical feedback, or the structural inclusion of additional independently-biased saturable-absorber sections, to induce the non-linearity required for neuron-like spiking behaviour. These approaches typically demand complex optical setups, precise control of experimental parameters (e.g. polarisation, optical power, wavelength detuning), sub-lasing threshold operation, and fine-tuned stabilisation, overall limiting their practical application and deployment.

This study introduces the first VCSEL system that intrinsically operates as a compact, single-section, free-running, coherent light-emitting (spiking) neuron. We refer to this device, which by design is a three-junction VCSEL, as a Neuron (Vertical-Cavity) Surface-Emitting Laser, or NeuronSEL. Unusual for these structures, the NeuronSEL exhibits a nonlinear S-shaped NDR region in both its Voltage-Current (V-I) and Light-Current (L-I) curves. This memristor-like NDR enables the NeuronSEL to produce self-sustained photonic–electronic neuronal features such as deterministic and tonic spike firing, refractoriness, threshold- and integrate-and-fire. This device, without external modulation or multi-section structures and biasing, acts on its own as an optical spiking source, delivering an output power of up to 2~mW at room temperature without thermal management. The NeuronSEL integrates in a single compact structure the beneficial features of a multi-junction VCSEL with the novel exploitation of NDR and excitability for neuronal functionality. This work also demonstrates the ability of the NeuronSEL to deliver neuromorphic processing functions, such as coincidence detection and XOR operation, realising an optical spiking neuron suitable for non-linear processing. Further, we demonstrate that the optical spiking emission from NeuronSELs can be seamless coupled with event-based camera systems, highlighting their potentials for novel neuromorphic photonic sensing platforms. Finally, we also showcase the excellent prospects for scalability of the NeuronSEL platform by emulating and studying a neuromorphic photonic SNN system formed by an array of 20 NeuronSELs able to perform dataset classification.

\section{Results}\label{sec:Results}
\subsection{NeuronSEL spiking: Static analysis}\label{Statics_results}

\begin{figure}[h]
\centering
\includegraphics[width=0.8\textwidth]{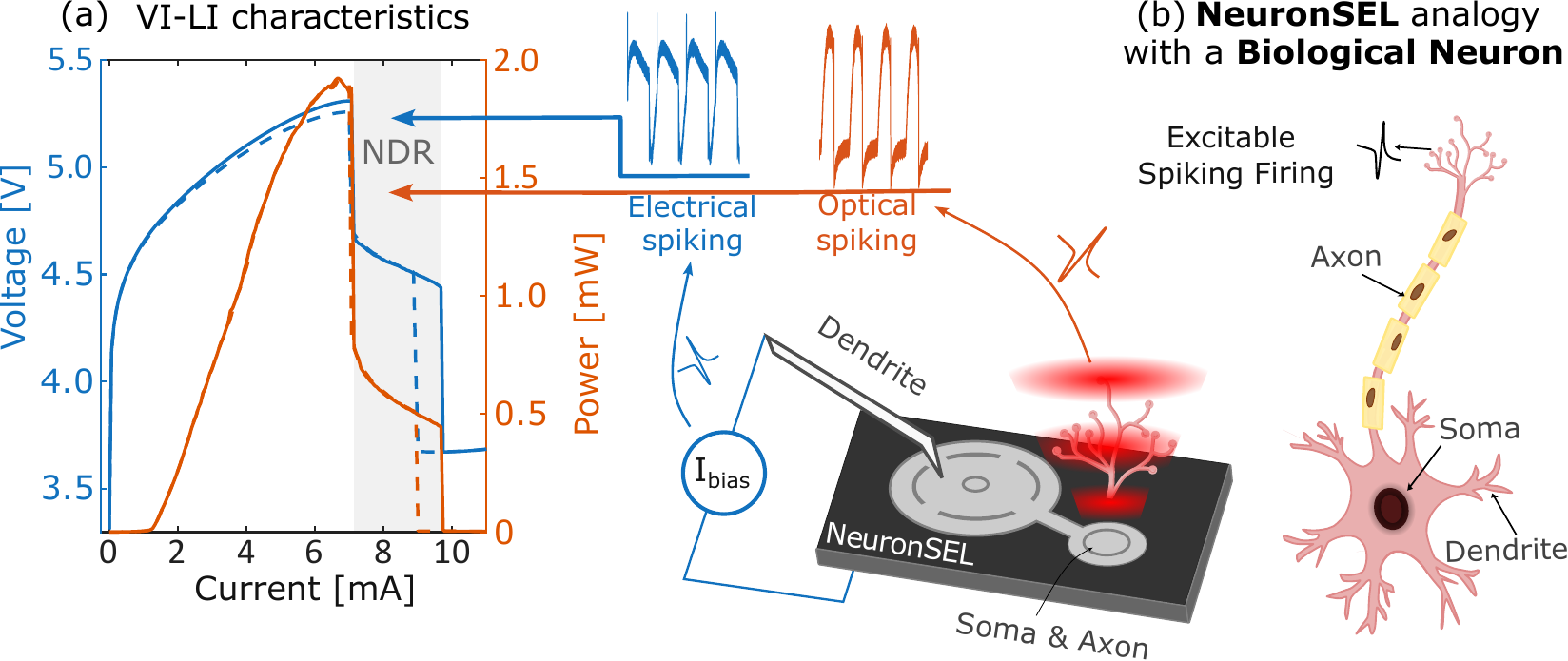}
\caption{(a) 
The Voltage–Current (V–I, blue) and Light–Current (L–I, orange) characteristics of the device highlight the Negative Differential Resistance (NDR) region, where electrical and optical spiking occurs. Continuous (dashed) curves correspond to the forward (reverse) bias sweep.
(b) Schematic representation of the analogy between the NeuronSEL and a biological neuron.
}\label{fig:Fig1_overview}
\end{figure}

Fig~\ref{fig:Fig1_overview} presents the nonlinear V-I and L-I characteristics of the free-running NeuronSEL, alongside the experimental setup and device structure (full details in Methods Section~\ref{Methods}). The device exhibits a threshold current of 1.3~mA and a turn-on voltage of 4.5~V (measured at room temperature without thermal management). The NeuronSEL exhibits an NDR region, where an increase in current results in an abrupt voltage drop (blue line) across the device. The increase in current initially produces a linear response in emitted optical power (orange line), until the device enters the NDR region, where it drops abruptly, before subsequently fully extinguishing the optical emission. 

\begin{figure}[h]
\centering
\includegraphics[width=0.9\textwidth]{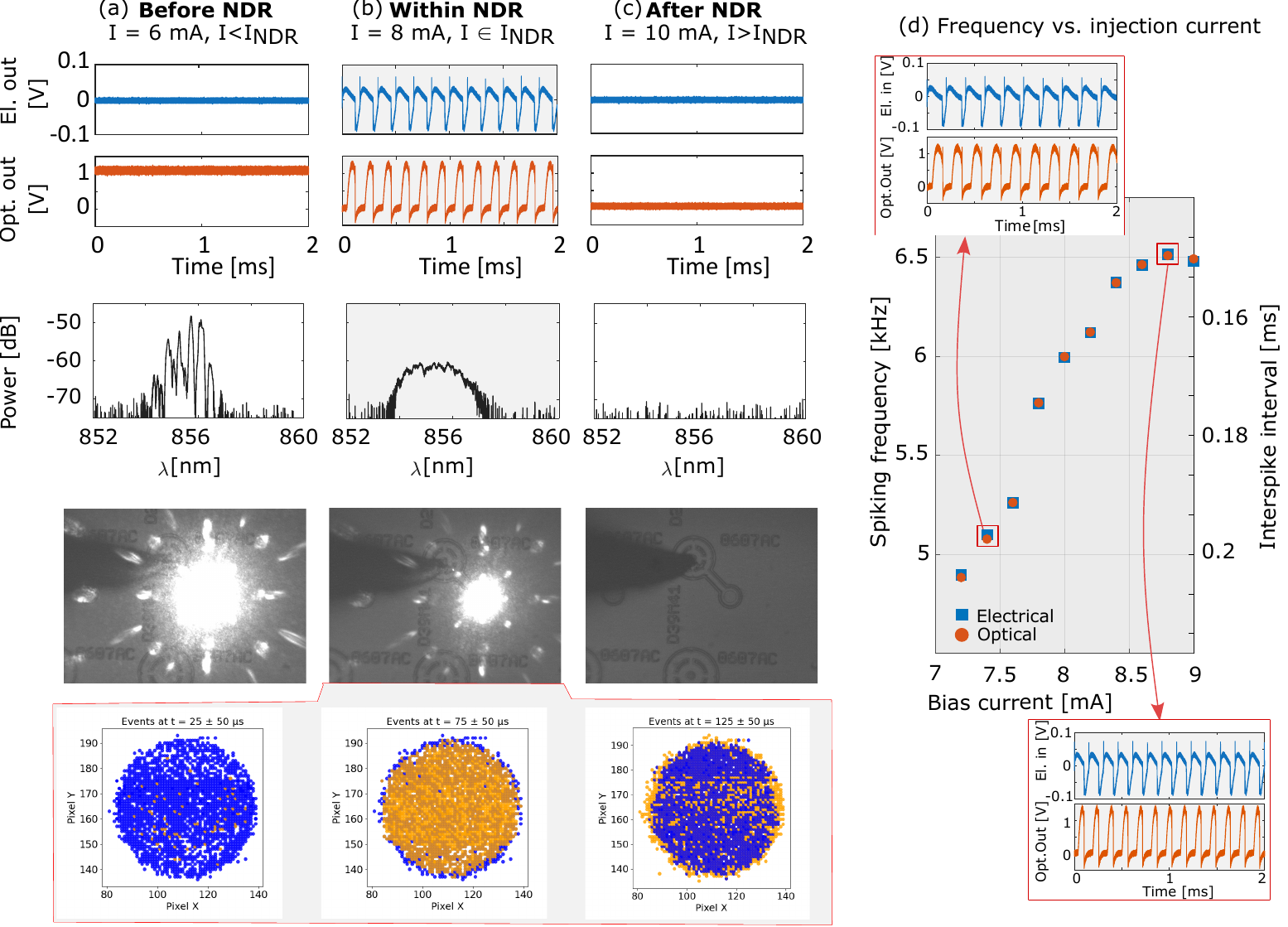}
\caption{NeuronSEL behaviour at different injection current values, $I$. (a) before the NDR $I<I_{NDR}$: 6 mA; (b) within the NDR  $I\in I_{NDR}$: 8 mA; (c) after the $I>I_{NDR}$; 10 mA. The top panels show the temporal traces for the electrical (blue) and optical (orange) output signals of the NeuronSEL. The middle panels display the corresponding optical spectra. The lower panels are captures of the NeuronSEL taken by a CMOS camera. The bottom panels show, at an injection current of 8 mA (b), the emission of the NeuronSEL captured using an event-based camera. Three slices of event-based camera data reveal a blinking (spiking) response, indicated by the changing polarity of events (blue-ON, orange-OFF). (d) Spiking frequency (in kHz) and interspike interval (in ms) for the electrical (blue squares) and optical (orange circles) spiking responses, at injection currents within the NDR.}\label{fig:Fig2_TS_spectra}
\end{figure}

Figs.~\ref{fig:Fig2_TS_spectra}~(a-c) illustrate the measured electrical, optical, and spectral responses of the NeuronSEL, for three different operation points (full spectral sweeps for all bias currents are presented in the Supplementary Material, Fig. S1).
Initially, operating below the NDR region ($I<I_{NDR}$) with $I=6$ mA (Fig.~\ref{fig:Fig2_TS_spectra}~(a)), the NeuronSEL follows the linear V-I and L–I regime, producing stable continuous-wave (CW) multi-transverse mode light emission centred around 856 nm.
Increasing the current to within the NDR region ($I\in I_{NDR}$, Fig.~\ref{fig:Fig2_TS_spectra}~(b) at $I=8$~mA, the system enters a periodic, self-sustained, neuron-like spiking regime. Spike-firing responses occur simultaneously in the optical emission and electrical voltage across the NeuronSEL, with positive and negative polarity, respectively. Fig.~\ref{fig:Fig2_TS_spectra}~(b)) shows that the optical spectrum broadens as the multi-transverse mode lasing emission of the NeuronSEL rapidly transitions (faster than the temporal resolution of the optical spectrum analyser) between on and off states in the spiking regime, creating on average the wide spectral profile.
Finally, when above the NDR region ($I>I_{NDR}$, Fig.~\ref{fig:Fig2_TS_spectra}~(c)) at $I=10$ mA, the voltage across the NeuronSEL drops abruptly and lasing ceases. Fig.~\ref{fig:Fig2_TS_spectra}~(d) shows that within the NDR region, the NeuronSEL exhibits electrical and optical spiking rates ranging from 4.9 to 6.5~kHz, with higher spiking frequencies achieved at larger currents. This highlights the system’s rate coding capability whereby stronger inputs are translated into faster spiking responses. Linked to the speed of the excitable (spike) trajectory, this behaviour is well understood in devices such as RTDs~\cite{Donati2025}. Discussed further in Methods Section \ref{Device_character}, we believe that the S-shape NDR in the NeuronSEL arises from thermally-activated alignment of electronic states in the cascaded active region structure. 

The emission from the NeuronSEL in each operating point was captured using two cameras: a standard CMOS camera, and an event-based vision sensor. Captures from the CMOS camera reveal high levels of light emission when the NeuronSEL operates in the linear regime below the NDR. However, when the device enters the NDR region, the detected light intensity by the CMOS camera decreases as the device begins to spike and alternate between on and off emission. Since the spike firing rate is higher than the frame-rate of the CMOS camera, the latter provides simply a reduced average intensity, and the NeuronSEL’s spiking optical signals couldn’t be temporally-resolved. Finally, when operating beyond the NDR region, the CMOS camera does not observe any light emission. The optical spiking signals emitted by the NeuronSEL when biased in the NDR region were however, successfully temporally-resolved using the event-based vision sensor (Prophesee EVK3), as seen in the three 2D temporal slices provided in Fig.~\ref{fig:Fig2_TS_spectra}. These plots show the sensor producing events (changes of detected light intensity) in response to the NeuronSEL’s circular beam. The three plots show an alternating change of net polarity, from positive (blue) events to negative (orange) events, corresponding to the spikes from the NeuronSEL (full reconstruction of the measured event-based signal is provided in the Supplementary Information, Fig.~S3). The resulting event-based optical link, seamlessly couples the native NeuronSEL to the event-based sensor, emphasising the high potential of both platforms being combined for future practical neuromorphic photonic systems and sensors.

\subsection{NeuronSEL spiking: Dynamic analysis}\label{sec:Dynamics_results}

\begin{figure}[h]
\centering
\includegraphics[width=0.9\textwidth]{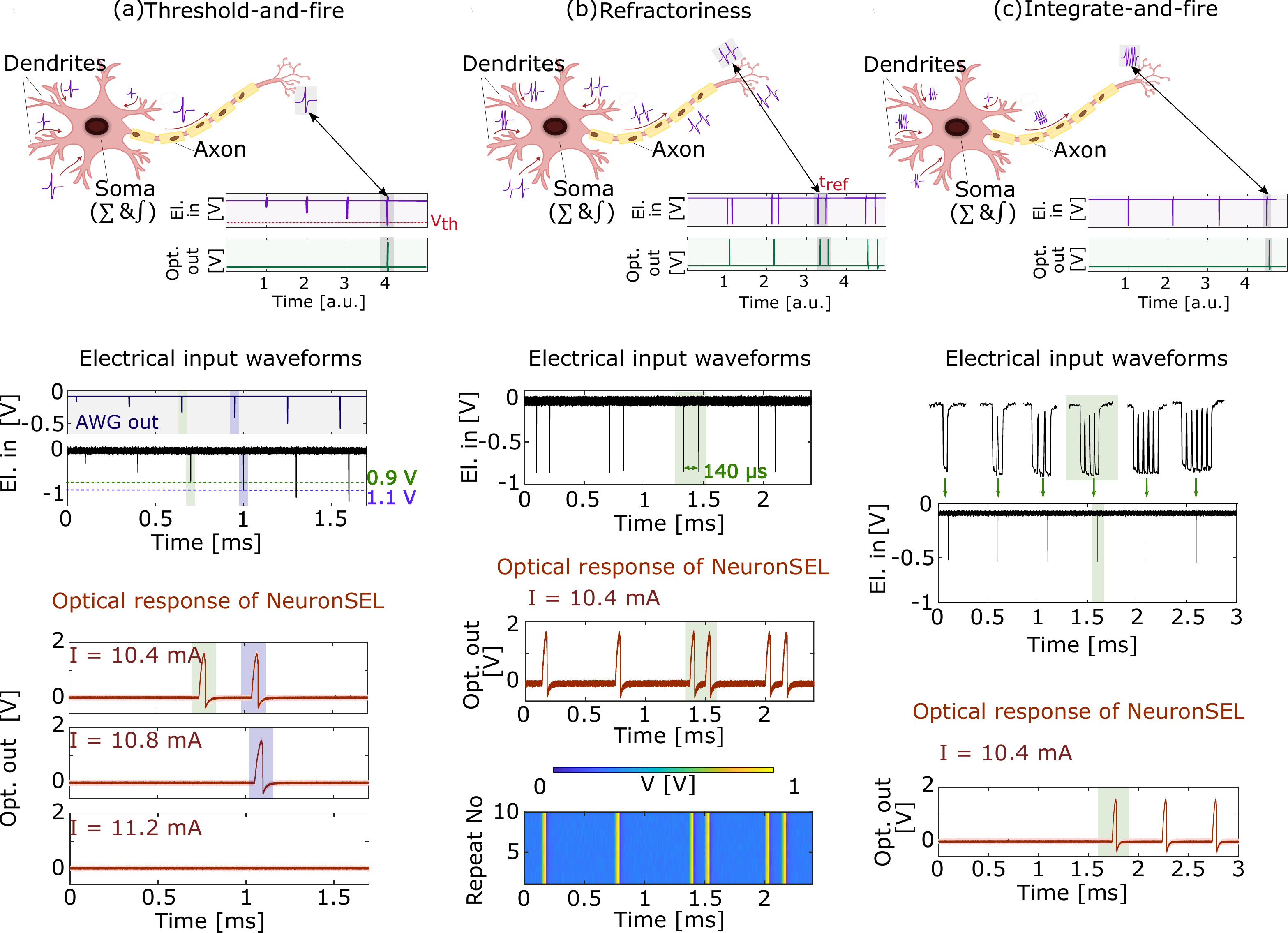}
\caption{Characterisation of key neuronal features using fast electrical voltage pulses.
    {(a) Threshold-and-fire.} 
    Top: Electrical input waveforms from the the AWG (dark blue) consisting of 10~ns pulses with increasing amplitudes from -0.1~V to -0.6~V with period of 300~$\mu$s, and amplified electrical input (black) highlighting the activation thresholds. 
    Bottom: Optical output traces for three bias currents above the NDR, $I = 10.4~\mathrm{mA}$, $I = 10.8~\mathrm{mA}$ and $I = 11.2~\mathrm{mA}$. 
    Light red lines show individual temporal traces, and dark red line indicates their average. 
    {(b) Refractoriness.} 
    Top: Amplified electrical input signal consisting of pairs of -0.9~V pulses of 10~ns width, with increasing separations from 110~$\mu$s to 150~$\mu$s in 10~$\mu$s steps, at a period of 500~$\mu$s. 
    Bottom: Optical response with the refractory period highlighted in green, with a colourmap of the optical responses over 10 repeated cycles.
    {(c) Integrate-and-fire.} 
    Top: Amplified electrical input consisting of bursts of 10~ns pulses with 3~ns separations, each at an amplitude of -0.5~V. 
    Bottom: Optical responses of the NeuronSEL highlighting the successful integration of a quadruplet burst.}\label{fig:Fig3_dynamics}
\end{figure}

Dynamic analysis of the neuromorphic responses in the NeuronSEL are also performed (Fig.~\ref{fig:Fig3_dynamics}~(a–c)), demonstrating deterministic triggering of spike firing regimes as well as key neuronal features such as threshold-and-fire, integrate-and-fire, and refractoriness. In these analyses, the device is biased with a constant current above the NDR region and subject to fast negative electrical voltage pulses.
\newpage

\textbf{Threshold-and-fire}

Fig.~\ref{fig:Fig3_dynamics}(a) demonstrates the existence of an input activation threshold, $V_{th}$, in the NeuronSEL beyond which spikes are deterministically produced.
The input waveform (displayed in dark blue), consists of a train of six 10~ns-wide pulses separated by 300~$\mu$s, with amplitudes ranging from -0.1~V to -0.6~V, which after amplification (plotted in black) are put across the device.
The NeuronSEL’s optical emission is measured for three bias currents above the NDR region. The optical responses overlay 10 repetitions with the average input plotted in dark red. At $I = 10.4$~mA, inputs of -0.9~V and -1.1~V deterministically trigger optical spikes. As the current increases to $I = 10.8$~mA, only the -1.1~V input triggers a spike event, with lower inputs no longer meeting the spiking threshold, similar to neuronal threshold-and-fire operation. For $I = 11.2$~mA, no spikes are triggered as the device operates well beyond the NDR region. Interestingly, large amplitude inputs (-1.3~V and -1.4~V) fail to produce spikes as the device transits into a stable state beyond the excitable regime, creating a dual threshold system. Beyond achieving deterministic firing with fast (ns-rate) voltage pulses, current-controlled spike firing is also demonstrated (see results in Supplementary Information, Fig. S2), permitting multiple, versatile operational modes. 

\textbf{Refractoriness}

Refractoriness, the characteristic time during which neurons cannot be re-excited after spiking, is also demonstrated in the NeuronSEL (Fig.~\ref{fig:Fig3_dynamics}~(b)). The device (biased at $I = 10.4$~mA) is subject to electrical input pairs (black signal) of -0.9~V pulses with 10~ns width and increasing separations, ranging from 100~$\mu$s to 160~$\mu$s, every 500~$\mu$s. The NeuronSEL’s optical response (red signal) shows that a second spike is only triggered when the inter-pulse separation is equal to or greater than 140~$\mu$s, confirming the presence of a refractory period analogous to biological neurons. The colourmap in Fig.~\ref{fig:Fig3_dynamics}~(b), displaying the optical output of the NeuronSEL in response to 10 repetitions of the electrical input signal, confirms the consistency of the measured refractory period in the system.

\textbf{Integrate-and-fire}

To evaluate the integrate-and-fire capability, the NeuronSEL was biased at $I = 10.4$~mA and subject to bursts of sub-threshold electrical inputs.
Illustrated in black in Fig.~\ref{fig:Fig3_dynamics}~(c), each consecutive burst is composed of an increasing number of 10~ns pulses, with 3~ns inter-pulse separation, pulse amplitude of -0.5~V and a separation between consecutive bursts of 500~$\mu$s. The corresponding optical responses of the NeuronSEL (red signal) show successful integration of sub-threshold inputs. Specifically, Fig.~\ref{fig:Fig3_dynamics}~(c) shows that when four or more pulses accumulate, the total combined input surpasses the activation threshold, triggering an optical spike. This demonstrates the operation of the NeuronSEL as a leaky integrate-and-fire neuron model, an important result given the neuronal model's wide spread use in neuromorphic processing~\cite{brunner2025roadmap,robertson2020ultrafast}.

\subsection{NeuronSEL functional tasks}\label{sec:Tasks_XOR_Coin_results}

Neuromorphic spike processing operations were also implemented with NeuronSELs. Firstly, we demonstrate XOR operation and temporal coincidence detection, before subsequently studying a scaled-up NeuronSEL-based photonic spiking neural network (pSNN) architecture, for performance in a dataset classification task.

\textbf{XOR task}\label{sec:Tasks_XOR}

\begin{figure}[h]
\centering
\includegraphics[width=\textwidth]{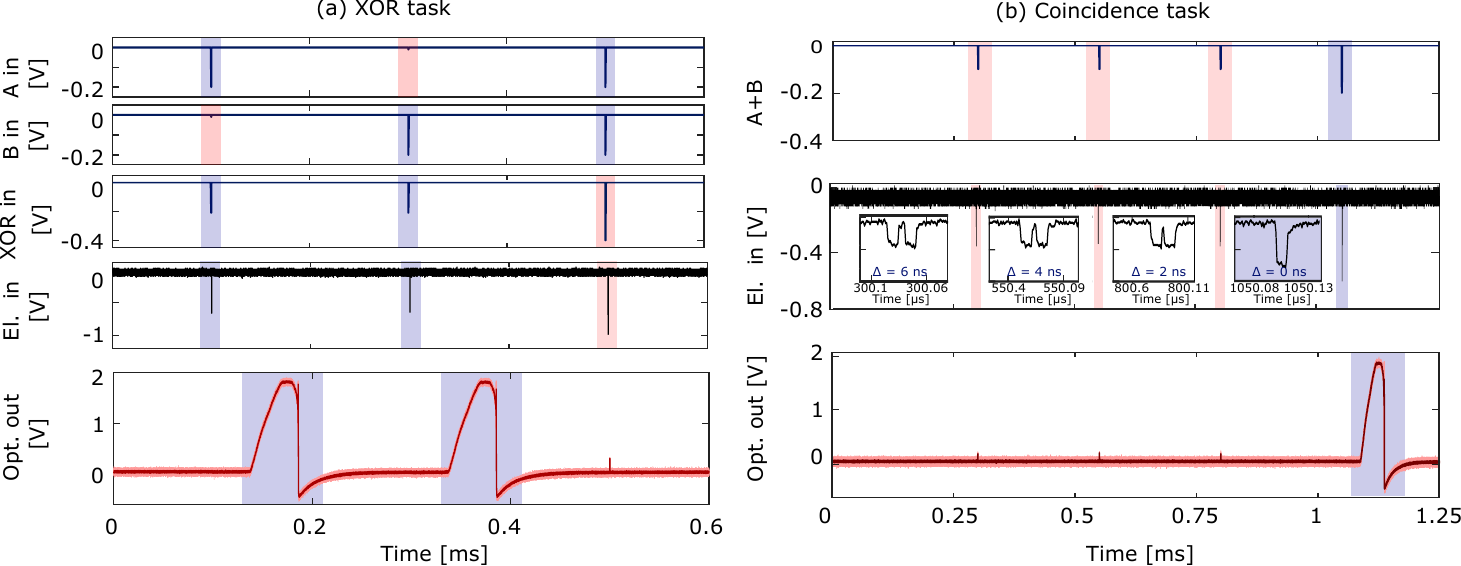}
\caption {Demonstrations of XOR and coincidence detection with the NeuronSEL.
    {(a) XOR task.} 
    Top: Pulse trains depicting Inputs A and B, respectively, and the combined and generated (black) electrical XOR input.
    Each input consists of -0.2~V pulses with 10~ns width placed at different temporal positions. 
    Bottom: Optical output response of the NeuronSEL to the XOR input at 10.4~mA.
    {(b) Coincidence task.} 
    Top: Pulse train depicting the combined Inputs A and B, and the genearted electrical input (black), where Input A is shifted around Input B with a decreasing temporal delay $\Delta$. Each input consists of -0.1~V pulses with widths of 10~ns. 
    Bottom: Optical response of the NeuronSEL at 10.4~mA, highlighting coincident detection at $\Delta$ = 0~ns.
    }\label{fig:Fig4_XOR_Coincidence}
\end{figure}

In this first demonstration, two input signals A and B (Fig.~\ref{fig:Fig4_XOR_Coincidence}~(a) top panels) introduce independent trains of 10~ns-long electrical pulses, which are combined and launched into the NeuronSEL. The total signal therefore consists of pulses of -0.2~V (when either input A or B act individually), and -0.4~V (when inputs A and B coincide). Following amplification, the total signal (Fig.~\ref{fig:Fig4_XOR_Coincidence}~(a), black) applies electrical pulses of -0.7~V and -1~V across the NeuronSEL. When biased above its NDR region (at $I = 10.4$~mA), the device outputs an optical spiking response (red signal) that performs an XOR operation. Individual inputs from A or B elicit optical spikes (logical TRUE state), while simultaneous inputs from A and B suppress the spiking response (logical FALSE). Uniquely to the NeuronSEL, the system can exploit its double-threshold characteristic around the NDR boundaries to suppress spike activation for combined inputs. The intrinsic XOR functionality in the NeuronSEL offers exciting prospects for the creation of novel non-linear neuromorphic processing units, with complexity and performance above and beyond simple summing devices. 
\newpage
\textbf{Temporal coincidence detection}

In this second demonstration, inputs A and B (Fig.~\ref{fig:Fig4_XOR_Coincidence}~(b) top panels) introduce 10~ns-wide sub-threshold pulses of -0.1~V. The A (blue) and B (green) inputs are separated by different temporal delays $\Delta$, ranging from 6~ns to 0~ns in steps of 2~ns. The optical response of the NeuronSEL (Fig.~\ref{fig:Fig4_XOR_Coincidence}~(b), red trace) shows that only when the two pulses coincide in time ($\Delta$ = 0~ns), the combined input exceeds the system’s excitability threshold, resulting in the firing of a single optical spike. This realises a coincidence detection task analogous to a temporal AND operation. Notably, due to the inherent temporal leaky-integrate-and-fire functionality of the NeuronSEL (see Section~\ref{sec:Dynamics_results}), several temporal inputs with $\Delta$ of only a few ns, will likely be considered coincident, even without direct temporal overlap. 

\label{sec:pSNN_NeuronSEL}
\newpage
\textbf{Photonic Spiking Neural Network (pSNN) with uncoupled NeuronSEL array}

\begin{figure}[h]
\centering
\includegraphics[width=0.9\textwidth]{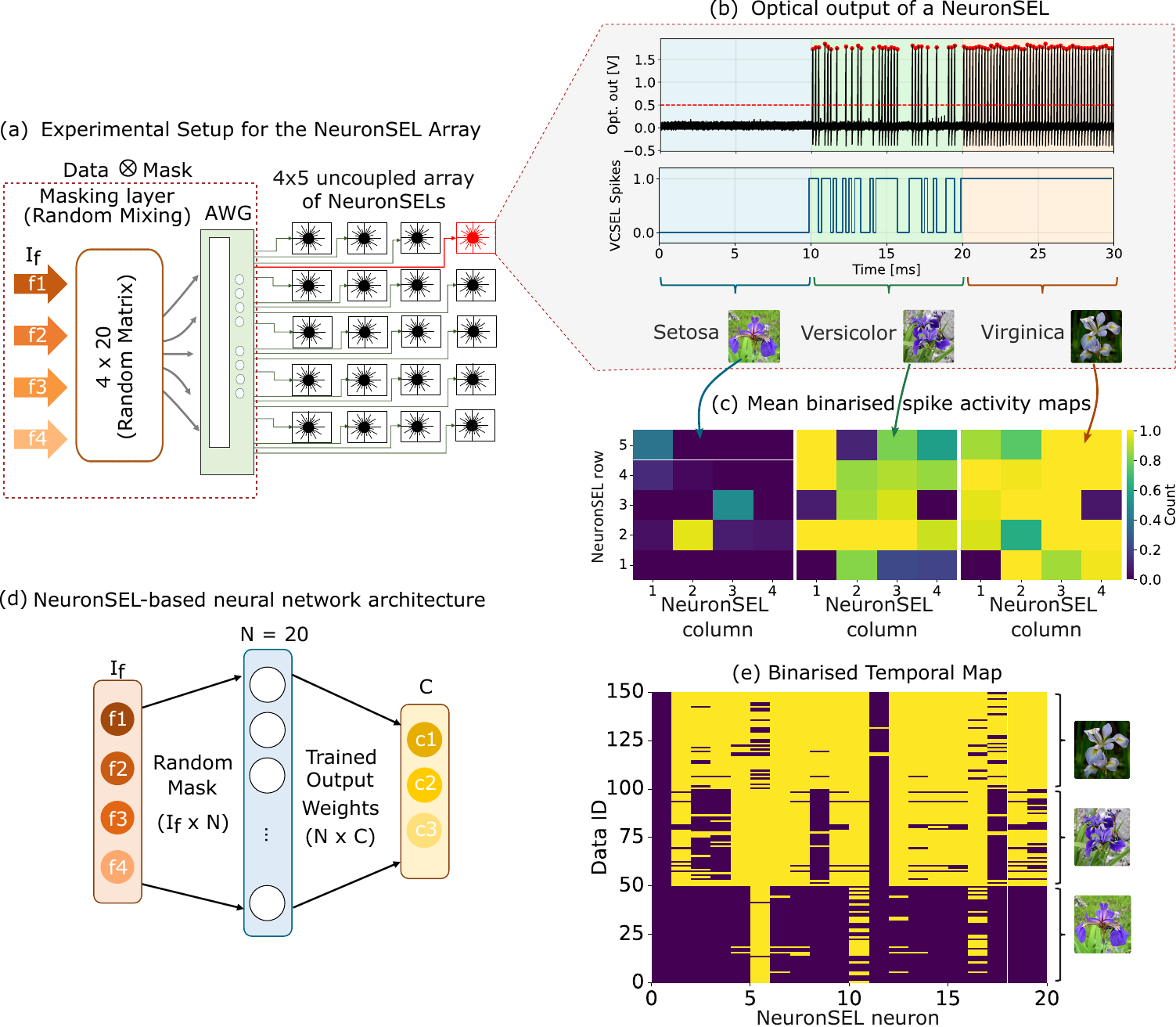}
\caption{Experimental NeuronSEL pSNN and analysis Iris flower classification performance. 
(a) Experimental setup of the proposed NeuronSEL pSNN. Input datapoint features ($I_f = 4$) are linearly mixed by applying a randomly generated masking matrix. The masked inputs are applied to each of the N = 20 NeuronSEL nodes in a $5 \times 4$ matrix array. (b) Optical output (black) of the highlighted NeuronSEL device, showing its response to the three classes: setosa (blue), versicolor (green), and virginica (orange). A red horizontal dashed line indicates the threshold ($th = 0.5$) used to detect spikes and create a binary output vector (1 for a spike, 0 otherwise). (c) Mean spike activity maps for each dataset class, highlighting evolving activity patterns between the different Iris flower classes. (d) Architecture of the proposed pSNN, comprising oft a masking input layer ($I_f = 4$), the NeuronSEL array (hidden layer, $N = 20$), and a trained readout layer ($C = 3$). (e) Temporal map showcasing the binary output vector for each datapoint.
}\label{fig:Fig_tasks}
\end{figure}

Using the Photonic Spiking Neural Network (pSNN) architecture described in Methods Section \ref{PSNN architecture}, we investigate a system formed by N~=~20 uncoupled NeuronSELs in a 5×4 arrayed structure (see Fig.~\ref{fig:Fig_tasks}~(a) and (d)), and evaluate its performance on the Iris flower dataset classification task~\cite{FISHER1936}. This NeuronSEL-based pSNN is emulated experimentally using a single device and time-multiplexing. The system is fed linearly-mixed features of Iris flower data, before creating 20 parallel optical spiking NeuronSEL nodes that are read-out independently and trained offline to perform the classification. The temporal optical spiking patterns generated by the 20 NeuronSEL network nodes are measured using a free-space photodetector (Fig.~\ref{fig:Fig_tasks}~(b), black trace) and binarised into a vector (blue trace). The mean spike activity maps (Fig.~\ref{fig:Fig_tasks}~(c)) display the normalised spike count for the time-multiplexed array of 20 NeuronSELs in response to each of the 3 Iris flower classes. These activity maps reveal clear differences in the measured optical spiking behaviour across the NeuronSEL array, prior to training. The temporal spiking response of each node is also shown in Fig.~\ref{fig:Fig_tasks}~(e), where characteristic patterns emerge for each of the 150 dataset points. Applying linear least squares regression to Fig.~\ref{fig:Fig_tasks}~(e), the investigated pSNN achieved accuracy of 94.83\%. This result was obtained using a small, simple, uncoupled network architecture (with only 20 NeuronSEL spiking nodes) trained with 10 datapoints per class. Least square training of the masked input signal showed a clear advantage to using the NeuronSEL array, achieving only 81\% accuracy when trained using 25 datapoints per class. Overall, this processing demonstration highlights the potential of NeuronSELs for scalable, practical and powerful pSNN architectures for future photonic-electronic neuromorphic processing platforms.  

\section{Discussion}\label{sec:conclusion}

In this work we introduce, fabricate, and experimentally demonstrate a new class of neuromorphic photonic device, the NeuronSEL. This device is based on a multi-junction VCSEL whose internal dynamics enables the achievement of highly-nonlinear V-I/L-I characteristics yielding negative differential resistance and excitability. The NeuronSEL defines a first-in-class free-running compact photonic spiking (laser) neuron, capable of generating coherent optical (and electrical) neuron-like spiking signals under solitary operation. This device eliminates the need for complex multi-section device structures (e.g. saturable absorbing regions) or precise optical injection protocols, while benefitting from the unique advantages of VCSEL technology (e.g. vertical-emission, low manufacturing costs, ease to integrate in large 2D arrays, high wall-plug efficiency). Deterministic triggering of spiking responses as well as key neuronal features (e.g. refractoriness, threshold- and integrate-and-fire) are also demonstrated in the NeuronSEL, as well as its ability to perform neuromorphic spike-processing tasks (e.g. XOR operation, coincidence detection). Finally, we also demonstrated system network scalability by emulating experimentally a pSNN, formed by an array of 20 uncoupled NeuronSELs, and its successful operation in the Iris flower dataset classification task. Overall, the NeuronSEL demonstrates a radically new neuromorphic photonic element, yielding NDR and optical and electrical spiking, that benefits from a compact structure, and vertical coherent light emission, enabling its application in neuromorphic photonic processing. Furthermore, its seamless compatibility with event-based technologies (e.g. dynamic vision sensors), offers excellent prospects for end-to-end neuromorphic systems, unifying optical spike generation and event-based sensing. Its inherent nonlinear dynamics, compatibility with array-level integration, and demonstrated functionality in SNN operation, position the NeuronSEL as a highly promising new technology for future large-scale spiking photonic processors, edge-AI sensors, and hybrid electronic–photonic neuromorphic architectures where ultrafast, energy-efficient inference becomes possible at the hardware level.

\section{Methods}\label{Methods}
\subsection{Device fabrication}\label{Device_fabrication}

\begin{figure}[h]
\centering
\includegraphics[width=0.9\textwidth]{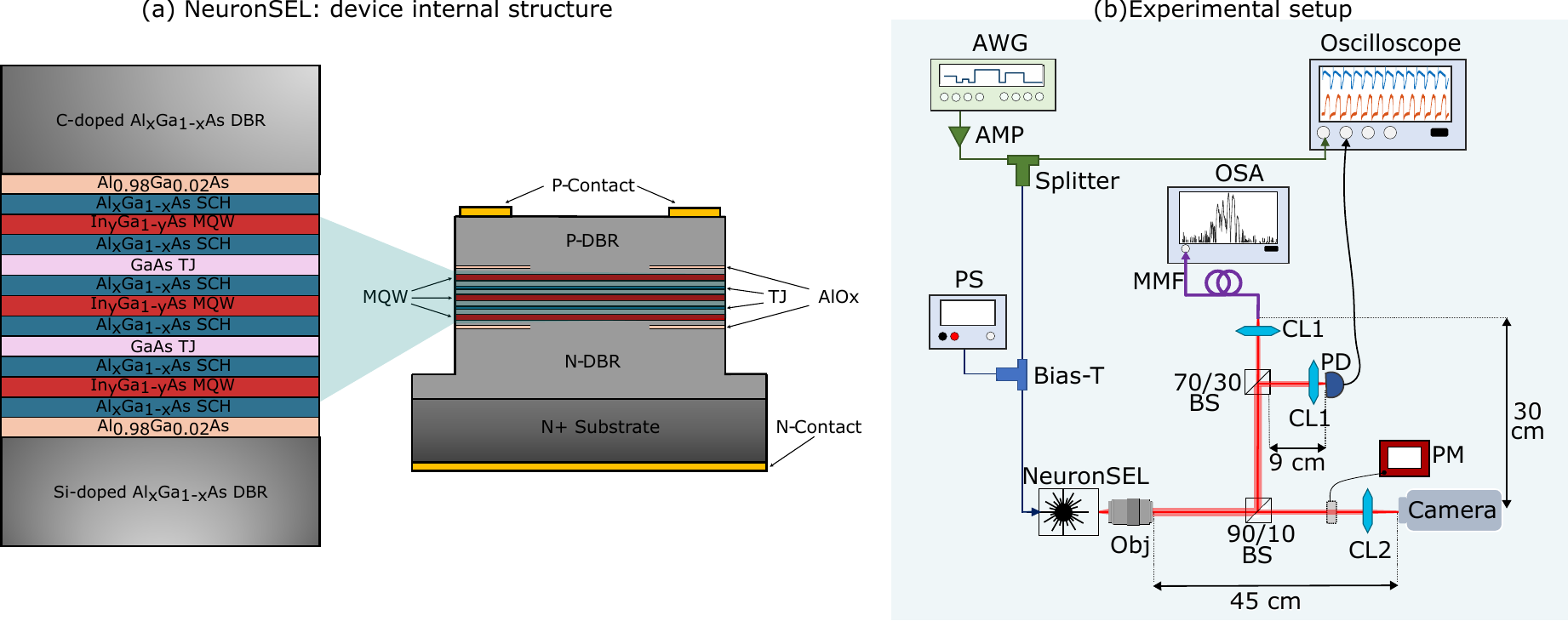}
\caption{(a) Internal structure of the NeuronSEL device.
(b) Experimental setup for NeuronSEL characterisation. AMP: Amplifier; Bias-T: Bias Tee; BS: Beam Splitter; MMF: Multimode Fibre; OSA: Optical Spectrum Analyser; PD: Photodetector; PM: Power Meter; PS: Sourcemeter.
}\label{fig:fig6}
\end{figure}

The NeuronSEL investigated in this work, whose internal architecture is detailed in Fig.~\ref{fig:fig6}~(a), is an oxide-confined multi-junction VCSEL designed for operation at around 850~nm. The active layers are InGaAs multi-quantum wells (MQWs) that are cascaded to form a triple-junction gain medium. The MQW active stages are separated by GaAs Esaki tunnel junctions doped with tellurium and carbon in the n-type and p-type layers, respectively. The optical thickness of the cavity is 5/2$\lambda$ and it is clad by an upper (p-type) and lower (n-type) distributed Bragg reflector (DBR) mirror formed by pairs of alternating Al\textsubscript{x}Ga\textsubscript{1-x}As layers. Two thin Al\textsubscript{0.98}Ga\textsubscript{0.02}As layers are included after the first DBR pair above and below the cavity which, after selective oxidation, provide electrical confinement of carriers and optical confinement of the mode. The devices fabricated from this epitaxial structure have a stripped-back process requiring one etch, a selective oxidation process, two metal depositions, and one anneal, allowing for a reduction in fabrication complexity and time-to-result. Further details on the fabrication method are provided in~\cite{baker2022vcsel}. For this study, devices of mesa diameters 36 to 49 $\mu$m were produced, with resulting oxide apertures of 1 to 14 $\mu$m. The NeuronSEL under test in this work corresponds to a 41 $\mu$m diameter mesa and a 6 $\mu$m oxide aperture.

\subsection{Device characteristics}\label{Device_character}

The NeuronSEL exhibits a nonlinear voltage–current (V–I) and power–current (L–I) response characterised by an NDR region, where an increase in current results in a voltage drop across the device, as shown in Fig.~\ref{fig:Fig1_overview}~(a). 
For injection currents up to $I=7.0$~mA, the NeuronSEL operates in a stable linear regime, with both the voltage and the emitted optical power increasing with bias current. As the bias current enters the NDR ($I~=~7.1~-~9.7$~mA), the linearity breaks, with both voltage and optical output abruptly decreasing despite increasing the current. This breakdown takes the form of an S-shape (current-controlled) NDR, similar to S-shape NDRs observed in memristive devices~\cite{Kumar2017}. We hypothesise that the NDR behaviour in the NeuronSEL is thermally activated and depends primarily on joule heating in the structure. When electrically driven, the NeuronSEL quickly reaches a critical current that produces a drop in voltage across the structure. This drop of voltage may be caused by a cascading increase in joule heating and current flow, that forms low resistance, high current channels through the device. Low resistance channels would also lead to the breakdown of optical emission as a high number of carriers in the cavity are lost to thermal effects and do not radiatively recombine within the QW regions. Within the NDR, the continuous thermal cycling creates swings of voltage and temporal windows of optical emission (spikes). The KHz cycling of the laser emission, added to the low bandwidth optical power meter used to measure the optical intensity from the NeuronSEL, produces the reduced average optical power measurement observed within the NDR. With operation at room temperature, without additional thermal management, we believe these thermal effects are the dominate contributor to the timescales observed in the NeuronSEL. Due to the presence of tunnel junctions, and the frequently exploited quantum-tunnelling pathway to N-shaped NDRs~\cite{Donati2025}, a voltage-controlled sweep of the structure was performed at an early stage. However, the voltage sweeps produced no visible NDR, indicating that tunnelling and band-bending are not likely the dominant mechanism for NDR formation in the NeuronSEL. Overall, the NeuronSEL’s characteristics realise for the first time to our knowledge, controllable and deterministic spiking responses, and controlled switching mechanisms in VCSELs through the manipulation of its inherent S-shape NDR.

\subsection{Experimental setup}\label{Experimental_setup}

The experimental setup depicted in Fig.~\ref{fig:fig6}~(b) was used to perform both the static and dynamic analyses of the NeuronSEL. A chip of NeuronSEL devices is mounted vertically where it is probed by tungsten needles and driven by an electrical input (DC and RF) via a bias tee (Bias-T). The DC bias is provided by a sourcemeter (PS, Keithley 2450) and the RF input is provided by an Arbitrary Waveform Generator (AWG, Keysight M8190A). The RF input is amplified (AMP, Exail DR-AN-20-MO) before being split with an electrical splitter to allow for the real-time observation of the voltage across the device. Optical emission from the NeuronSEL enters the free-space optical system where light is collected and collimated using a 10$\times$/0.3 air objective (Obj, Nikon CFI Plan Fluor). Collimated light is split by a 90/10 (R/T) beam splitter (BS), with 90\% of the light further divided by a second 70/30 (R/T) BS to enable simultaneous monitoring of the device’s optical output and optical spectrum, respectively. For spectral measurements, light is focused through a collimating lens (CL1, Thorlabs AC254-030-AB-ML, $f~=~30$~mm) and onto a multimode fibre (MMF), subsequently connected to an Optical Spectrum Analyser (OSA, Yokogawa AQ6370E). For measurements of the NeuronSEL’s optical output, light was again focused using a collimating lens (CL1) and detected with a free-space fast amplified photodetector (FPD610-FS-NIR). The output optical and electrical signals were recorded using a 16~GHz, 40~GSa/s real-time oscilloscope (OSC, R\&S RTP). The remaining 10\% of the emitted light was focused through a second collimation lens (CL2, Thorlabs AC254-100-AB-ML, $f~=~100$~mm) and projected onto a CMOS camera sensor (IDS, U3-3682XLE-NIR Rev.1.2 uEye+). The optical path lengths are indicated in the experimental setup diagram. For monitoring the total optical output power, a power meter (Thorlabs S120C \& PM100A) was placed in the camera arm of the setup. Simultaneous recordings of the optical and electrical responses were taken using the real-time oscilloscope, with optical spectra and output power measured for each injection current value. To capture event-based vision sensor data, a prophesee EVK3 camera was used in place of the CMOS camera sensor.

\subsection{Photonic Spiking Neural Network (pSNN) architecture}\label{PSNN architecture}

A photonic spiking neural network composed of $N$~=~20 nodes is implemented to demonstrate the scalability of the NeuronSEL. In this proof-of-concept demonstration the pSNN is created using a single NeuronSEL and time-multiplexing. A single NeuronSEL is driven consecutivly with multiple different input signals to create the uncoupled 5$\times$4 (20 node) NeuronSEL array. The network performance is assessed using the Iris flower dataset, a classification problem that consists of three classes (setosa, versicolor, and virginica), each containing 50 datapoints. Each datapoint is described by four features ($I_f$~=~4): petal length, petal width, sepal length, and sepal width. The input features are linearly mixed using a randomly generated mask of dimensions $I_f$ $\times$ N (4$\times$20). This random mask simulates the effect of propagating the incoming information through a scattering medium before it is injected into the NeuronSEL array. The linear mixing generates a distinct weighting for each of the NeuronSEL nodes. All datapoints are run sequentially with the weighted inputs forming trains of electrical pulses. Each electrical pulse is held at the weighted input value for 40~ns before returning to zero. An interval of 200~$\mu$s is used between datapoints to allow each neuron to relax (overcoming its inherent refractory period) before the next pulse is injected. For signal generation, the AWG used a sampling rate of 125~MSa/s. The optical spiking pattern generated by each node is directly measured before post-processing is performed offline. For each datapoint interval, a fixed threshold ($thr~=~0.5$) is applied to create a binary vector of the optical spikes. The binary vectors of all nodes are combined before training is applied to the data. Output layer weights are calculated using linear least square regression before the final readout and classification performance is analysed. Training is performed with 10 randomly selected datapoints from each class (30 datapoints in total) and testing is carried out on the remaining datapoints (40 datapoints per class, 120 in total).


\backmatter

\bmhead{Data Availability}
All data underpinning this publication are openly available from the University of Strathclyde KnowledgeBase at http://doi.org/10.15129/571dce29-f5c9-4d73-b321-26278fbebec3

\bmhead{Author Contributions}
M.D.G and J.R equal contribution. Idea and Concept: A.H, S.S, and J.R. Device Fabrication: J.B, C. P. A., S.S and P.M.S. Device Test: J.B, M.D.G, and J.R. System Design: J.R, M.D.G, X.P and A.H. System Measurement: M.D.G and J.R. Investigation and Analysis: M.D.G and J.R. Algorithms and training: M.D.G, J.R and D.O.N. Writing: M.D.G, J.R and J.B. Review and editing: All authors. 

\bmhead{Acknowledgements}
The authors acknowledge support from the UKRI Turing AI Acceleration Fellowships Programme (reference EP/V025198/1), EU Pathfinder Open project ‘SpikePro’ (Grant ID 101129904), UK Multidisciplinary Centre for Neuromorphic Computing (reference UKRI982), EPSRC Project ‘ProSensing’ (reference EP/Y030176/1), UKRI Strength in Places Fund (107134), The Royal Society (Grant RG/R2/232230), EPSRC Compound Semiconductor Manufacturing Hub for a Sustainable Future (reference EP/Z532848/1), and CS Underpinning Equipment Grant (reference EP/P030556/1) all provided essential resources for this study.

\bibliography{4_sn-bibliography}

\end{document}